\documentclass[12pt,epsf,amssymb]{article}

\usepackage{graphicx}
\usepackage{axodraw}
\usepackage{amsfonts}
\usepackage[usenames]{color}
\usepackage{amsmath}

\begin{document}
\baselineskip 0.6cm
\newcommand{\gsim}{ \mathop{}_{\textstyle \sim}^{\textstyle >} }
\newcommand{\lsim}{ \mathop{}_{\textstyle \sim}^{\textstyle <} }
\newcommand{\vev}[1]{ \left\langle {#1} \right\rangle }
\newcommand{\bra}[1]{ \langle {#1} | }
\newcommand{\ket}[1]{ | {#1} \rangle }
\newcommand{\Dsl}{\mbox{\ooalign{\hfil/\hfil\crcr$D$}}}
\newcommand{\nequiv}{\mbox{\ooalign{\hfil/\hfil\crcr$\equiv$}}}
\newcommand{\nsupset}{\mbox{\ooalign{\hfil/\hfil\crcr$\supset$}}}
\newcommand{\nni}{\mbox{\ooalign{\hfil/\hfil\crcr$\ni$}}}
\newcommand{\EV}{ {\rm eV} }
\newcommand{\KEV}{ {\rm keV} }
\newcommand{\MEV}{ {\rm MeV} }
\newcommand{\GEV}{ {\rm GeV} }
\newcommand{\TEV}{ {\rm TeV} }

\def\diag{\mathop{\rm diag}\nolimits}
\def\tr{\mathop{\rm tr}}

\def\Spin{\mathop{\rm Spin}}
\def\SO{\mathop{\rm SO}}
\def\O{\mathop{\rm O}}
\def\SU{\mathop{\rm SU}}
\def\U{\mathop{\rm U}}
\def\Sp{\mathop{\rm Sp}}
\def\SL{\mathop{\rm SL}}
\def\simgt{\mathrel{\lower2.5pt\vbox{\lineskip=0pt\baselineskip=0pt
           \hbox{$>$}\hbox{$\sim$}}}}
\def\simlt{\mathrel{\lower2.5pt\vbox{\lineskip=0pt\baselineskip=0pt
           \hbox{$<$}\hbox{$\sim$}}}}

\def\change#1#2{{\color{blue} #1}{\color{red} [#2]}\color{black}\hbox{}}


\begin{titlepage}

\begin{flushright}
UCB-PTH-05/17 \\
LBNL-57714
\end{flushright}

\vskip 2cm
\begin{center}
{\large \bf Density Perturbations and the Cosmological Constant\\
from Inflationary Landscapes} 

\vskip 1.2cm
Brian Feldstein, Lawrence J. Hall, and Taizan Watari.

\vskip 0.4cm
{\it Department of Physics and Lawrence Berkeley National Laboratory,

University of California, Berkeley, CA 94720} \\

\vskip 1.5cm

\abstract{An anthropic understanding of the cosmological constant
  requires that the vacuum energy at late time scans from one patch of
  the universe to another. If the vacuum energy during inflation also
  scans, the various patches of the universe acquire exponentially
  differing volumes.  In a generic landscape with slow-roll inflation, we find
  that this gives a steeply varying probability
  distribution for the normalization of the primordial density perturbations, 
  resulting in an exponentially small fraction 
  of observers measuring the COBE value of $10^{-5}$.  Inflationary landscapes should
 avoid this ``$\sigma$ problem", and we explore features that can allow them 
to do that.
  One possibility is that, prior to slow-roll inflation, the probability distribution
  for vacua is extremely sharply peaked, selecting essentially a
  single anthropically allowed vacuum. Such a selection could occur in
  theories of eternal inflation.  A second possibility is that the
  inflationary landscape has a special property: although scanning
  leads to patches with volumes that differ exponentially, the value of the
  density perturbation does not vary under this scanning. This second case
  is preferred over the first, partly because a flat inflaton potential can 
  result from anthropic selection, and partly because the anthropic selection of a 
  small cosmological constant is more successful.  }

\end{center}
\end{titlepage}


\section{Introduction}

The fundamental parameters of the standard model of particle physics 
and the standard Big-Bang cosmology are determined 
only from experiments and observations. One of the most 
important problems of physics is to provide a theoretical
understanding for the values of these parameters. 
Such ideas as unification, symmetry and naturalness have had 
partial success, bringing radiative corrections under control 
and reducing the number of independent parameters.  
The small non-zero cosmological constant (CC), however, still seems to defy 
any explanation from these considerations \cite{Weinberg-review}.

The anthropic principle --- that observed values of parameters must allow for 
the existence of observers --- sets the stage for one of the rare
successful explanations for why the CC
is so small compared with its natural order of magnitude.
It also predicts that the CC should be non-zero, 
and this may explain the observation that the expansion of the universe has
recently begun to accelerate. 
Suppose that the fundamental theory of physics possesses many 
vacua with different values of the CC. 
The various vacua are realized cosmologically in different patches of 
the universe; ours survives  
anthropic selection only because the CC
is sufficiently small to allow large scale structure and gravitationally 
bound systems to form \cite{WeinbergI}.
This argument sets an upper bound on the CC
\begin{equation}
\Lambda^4 \simlt  \left[ \rho_{\rm CDM} \delta^3 \right]_{rec}, 
\label{eq:WeinbergI}
\end{equation} 
where $\rho_{\rm CDM}$, the energy density of cold dark matter,
and $\delta$, the density perturbation for galactic sized modes, 
are evaluated at the epoch of recombination. 
The upper bound (\ref{eq:WeinbergI}) is only about a factor 
of $10^2$ higher than the observed value of the CC.
This is an enormous improvement over the naturally expected value,
which is  $10^{120}$ times larger than the observed value.

While there is no direct experimental evidence 
that the CC is determined by this mechanism, stringent anthropic
constraints on the values of 
the QED and QCD coupling constants \cite{anthropicalpha} also suggest  
that there are plenty of vacua on which cosmological selection acts; 
otherwise our existence would be a remarkable coincidence. 
Cosmological selection may also explain why we live 
in a vast homogeneous universe created by inflation.
Although fine-tuning of parameters is generically required to obtain
a sufficiently flat potential for slow-roll inflation 
\cite{eta2,eta3}, a vacuum with finely-tuned parameters that 
leads to successful inflation dominates the volume of the universe, 
giving an anthropic prediction for a flat potential \cite{Vilenkin95}. 
The field space of the underlying theory, containing lots of vacua 
with different values of various parameters \cite{Hawking,baby}, 
has recently been called the landscape \cite{string-landscape} and 
has been studied extensively, mainly in the context of string theory. 

Once we accept that there may be many vacua, realized in various patches 
of the universe, the notion of naturalness is replaced by probability. 
We assume that the probability, ${\cal P}$, of measuring a given value of 
a parameter is given by the fraction of observers in the universe 
who see that value.
This probability takes into account not only the density of vacuum states 
in the landscape, but also appropriate weight factors arising
from cosmological dynamics and selection, and can be decomposed 
as \cite{Vilenkin95}:
\begin{equation}
d {\cal P}(\xi) \propto 
  {\cal I}(\xi) {\cal V}(\xi) {\cal A}(\xi) d\xi,
\label{eq:Vilenkin-formula}
\end{equation} 
where $\xi$ denotes a collection of parameters of the low energy theory 
that vary from one vacuum to another. The
factors ${\cal I}$, ${\cal V}$ and ${\cal A}$ stand for the initial volume  
distribution prior to slow-roll inflation, the cosmological volume 
increase due to slow-roll inflation, and 
the ``anthropic factor", respectively. 
The first factor ${\cal I}(\xi)$ may come from a density of states, perhaps
calculated from the underlying statistics of vacua in string theory \cite{string-landscape}, and weighted, for example,
by some quantum creation process of the universe 
\cite{Vilenkin82,HH,Linde-wavefcn,HT,Cornell}.
The number of observers is also proportional to the volume factor 
${\cal V}(\xi)$, and it is the consequences of this very large factor 
that we explore in this paper.
The last factor ${\cal A}(\xi)$ includes all other weightings associated 
with the existence of observers. 

Rather than attempting to define the concept of an observer, 
we only consider a restricted set of patches of the universe
where the low energy effective theories,
cosmologies or environments are mildly perturbed about our own.\footnote{
Thus, we are not in a position to claim that certain collections of vacua,
combined with anthropic selection, uniquely  lead to the standard model and 
to the standard cosmology ({\it c.f.} \cite{A}) with parameters that can be determined.
Based on the restricted set, however, we can discuss necessary properties which must be satisfied by a landscape
 of vacua, along with the relevant cosmological dynamics, so that the cosmological constant and the density perturbations may be predicted 
correctly. The resulting conditions need not be sufficient, however.} 
After inflation and reheating, observers are created at a certain rate per unit volume, and
for a fixed period of time that ends when stars have burned up 
all of their available fuel.  
The factor ${\cal A}(\xi)$ is proportional to the number of observers produced 
per unit physical volume, and depends, for example, on the number density 
of acceptable galaxies formed.

As the CC approaches the upper bound (\ref{eq:WeinbergI}), a smaller 
and smaller fraction
of baryons form galaxies  \cite{Efstathiou,WeinbergII}, causing the anthropic
factor ${\cal A}$ to shrink; fewer observers are expected 
to see the value of such a large CC. 
The authors of reference \cite{WeinbergII} assumed that the only 
relevant quantity that scans independently is the CC, 
i.e., $\xi = \Lambda^4$, and that ${\cal I}(\xi){\cal V}(\xi)$ is 
$\Lambda^4$-independent. In this case, with every small value 
of $\Lambda^4$ represented equally in the density of states, 
they concluded that 5--10 \% of observers in the universe, 
rather than a fraction $10^{-2}$, would see a CC 
smaller than the value observed by us. 
This is a remarkable success.
It may be that the only parameter of nature that is significantly
scanned in cosmology is the CC itself,  so that this
result justifies attempts to
understand fundamental particle physics while ignoring the
CC problem, and we have nothing to add.

However, if the CC scans, then why not other parameters?
In this case one must study whether the scanning of multiple parameters
can maintain a successful understanding of the CC.
Suppose that the underlying theory possesses $N$ parameters which scan.
Some number $n$ of the standard-model parameters, such as the gauge couplings, 
have allowed anthropic windows in ${\cal A}$ that are so narrow 
\cite{anthropicalpha} that anthropic selection will determine $n$ combinations 
of the scanning parameters, leaving $N_s = N-n$ freely scanning. 

Let $V(\phi)$ be the classical potential energy density of the universe,
with $\phi$ representing all the scalar fields of the theory, 
including the inflaton. 
This potential contains many terms, each depending on a fundamental parameter 
and each typically much larger than the CC.
Since $\Lambda^4 =  V(\vev{\phi})$,
an anthropic understanding of the CC implies that some parameter(s) of 
$V(\phi)$ must scan, allowing cancellations between the various terms.
The special case of $N_s = 1$ allows the CC to scan but nothing else. 
What happens in the more general case of $N_s >1$? 
Since inflation is governed by $V(\phi)$, one now expects that the number 
of e-foldings of inflation, $N_e$, will also scan, leading to a crucial 
effect on the number of observers \cite{Vilenkin95,GL}.\footnote{Even for the case of $N_s = 1$, 
with one continuous parameter controling both the CC and the inflaton 
potential, one might wonder whether the inflation volume factor ${\cal V}$ 
could be so important that the ``a priori'' distribution ${\cal IV}$ of 
\cite{WeinbergI,WeinbergII} is no longer flat in $\Lambda^4$,
 weakening the anthropic understanding of the CC. 
However, providing $N_e$ is a mild function of the parameter and 
$N_e \simlt 10^{120}$, the distribution  ${\cal IV}$ is sufficiently flat 
for the small values of $\Lambda^4$ that are of interest.}
When all else is held fixed, the number of observers is proportional to 
the total volume in which they live,
so that
\begin{equation}
{\cal V}(\xi) \propto  e^{3N_e(\xi)},
\label{eq:Volume}
\end{equation}
where $3N_e(\xi) \simgt {\cal O}(100)$ varies as a
function of parameters. 
If $N_s > 1$, allowing the parameters of inflation to scan, 
there is no doubt that the volume factor ${\cal V}$ is likely 
to be a decisive part of the 
probability calculation. 
If the universe undergoes eternal inflation \cite{Vilenkin-eternal,
Linde-eternal,Susskind}, the corresponding volume factor may become 
even more important \cite{Vilenkin95,GL}.

The anthropic selection that results from maximizing $N_e(\xi)$ 
will have important consequences for
the observed primordial density perturbations, assuming they are
generated from the quantum fluctuations of a field during
inflation. The amplitude for some specific mode will have a
Gaussian probability distribution, proportional to
$e^{-\delta^2/\sigma^2}$, where $\sigma$, the scale of the density
perturbations, is computed in terms of the
parameters of the inflaton potential.  Unless specified otherwise, $\delta$ and
$\sigma$ will always refer to their values at the time the perturbations re-enter the horizon. 
The anthropically allowed window  for $\sigma$ is quite broad $10^{-6} \simlt \sigma \simlt  10^{-4}$ \cite{RT}, 
with the number of observers, and hence the
factor $\cal{A}$, falling off rapidly with $\sigma$ outside 
the window.\footnote{For larger $\sigma$, the density perturbations go 
non-linear when the average energy density of the universe was higher, 
so that the resulting structures are too crowded to guarantee 
a stable environment for life to evolve.
Below the lower boundary of $10^{-6}$, the majority of overdense regions 
are not able to cool quickly enough to form fragmented, structured galaxies.
On the other hand, for $10^{-6} \simlt \sigma \simlt 10^{-4}$ the dependence of 
$\cal{A}$ on $\sigma$ should be relatively mild.}  
If $N_s >1$, allowing scanning in the inflaton sector, not only is each patch
inflated by a different volume factor but the value of $\sigma$
differs in each patch. The question immediately arises as to whether
the patches with large $N_e$ typically have $\sigma$ close to the observed 
value of $\sim 10^{-5}$, or whether 
the volume factor (\ref{eq:Volume}) strongly favors other values.\footnote{
 From (1) it appears that a higher value of $\sigma$ is preferred
since it allows a higher value for the CC, weakening the 
success of the anthropic argument for the CC \cite{Wise}.
However, the selection of $\sigma$ is likely to be strongly dominated 
by the exponential appearing in the volume factor (\ref{eq:Volume}).}
In the generic case of $N_s >1$, it is necessary to determine the combined 
probability distribution for both $\sigma$ and $\Lambda^4$, 
and the success of the anthropic arguments of \cite{WeinbergI,WeinbergII} 
are far from guaranteed. In this paper we study whether such a probability 
distribution permits an anthropic understanding of the density perturbation 
as well as the CC, and, if so, in which landscapes.

In section \ref{sec:ensemble} we study simple field-theory models of
landscapes where the parameters of inflation are scanned cosmologically,
calculating the volume factor as a function of $\sigma$. Its dependence
is so steep that the fraction of observers measuring $\sigma \sim 10^{-5}$,
in the center of the anthropic window, is exponentially small. 
We argue that this is a generic problem of landscapes where the parameters 
of inflation models are scanned. 
While such scanning offers the hope of understanding the flatness of the inflaton potential, it 
leads to a ``$\sigma$ problem" of proportions at least as overwhelming as that of the 
CC.
Hence, we proceed to investigate whether landscapes with certain properties 
can overcome this $\sigma$ problem.
In section \ref{sec:eternal}, we describe a class of landscapes where 
initial conditions prior to slow roll inflation solve the $\sigma$ problem, and argue
 that eternal inflation may provide a mechanism to achieve this.  We will show that such models
can only give a probability between about $10^{-2}$ and $10^{-9}$ 
that we see a CC as small as we do, however. 
Section \ref{sec:restricted} is devoted to another class of landscapes,
where a restricted scanning of parameters can avoid 
the volume factor from being exponentially sensitive to $\sigma$.
In these cases, mild distributions for both $\sigma$ and $\Lambda^4$ allow 
$\sigma$ to naturally take a value $10^{-5}$ in the center of the anthropic 
window, and also allow an improved understanding of the observed CC, 
due to the anthropic factor ${\cal A}$ of \cite{WeinbergII}. 
Conclusions are drawn in section 5.

\section{Scanning in Models of Slow-Roll Inflation}
\label{sec:ensemble}
\subsection{One Parameter Model---Chaotic Inflation Ensemble}
\label{ssec:chaotic}

Let us first consider a simple field theory model of a landscape.\footnote{
In string-theory landscapes, space-time 
is not necessarily four-dimensional, and moreover, the Planck scale of the 
D=4 effective theory need not remain fixed relative to the string scale. 
All the compactified configurations that eventually ({\it c.f.} \cite{KR}) 
lead to decompactification in four dimensions, for example, through inflation involving 
D3-branes, are treated in our framework however. 
We do not consider a cosmological scan of the Planck scale 
in this article, because we think of it as the unit of all measurements: 
Any measurement is a  comparison between two observables of the same 
dimensionality, and we take the local value of the Planck scale 
as the basis for comparison.   We do this because
i) the string scale is not directly observable for the moment, 
and ii) because it is the ratio of the Hubble parameter or the
W-boson mass to the local value of the Planck scale, 
rather than to the string scale, that matters in physics within the sub-universes.
Thus, when we say that an inflaton mass parameter $m$ is scanned 
in a landscape, 
it can be interpretted in the application to string landscapes that 
a distribution of $m/M_{\rm pl}$ is obtained as a result of scanning 
of both/either $m$ and/or $M_{\rm pl}$.} 
The scalar potential $V(\phi)$ on the landscape may contain
many hills and valleys; some regions provide 
slow-roll inflation with sufficient e-folding numbers, while others do not. 
There will be many local minima; some contain the standard model 
as the low-energy effective theory, others do not. 
We are interested only in the inflationary regions leading 
to the standard-model minima. We expand the potential of these 
inflationary regions around the local minima, and approximate them by 
$V(\phi)= m^2 \phi^2$, where $m$ is a coefficient that has 
a different value for each region.  
We thus have an ensemble of chaotic inflation models. 
We assume that the quadratic approximation is valid even for 
$\phi$ significantly larger than $M_{\rm pl}$; those local regions that 
do not satisfy this criterion are discarded from the ensemble 
since they do not give sufficient inflation. 
This model landscape will illustrate how we obtain the probability 
distribution on $\sigma$, and why it depends exponentially on $\sigma$.

Initially, the universe is assumed to have local patches 
scanning over the inflationary regions in different parts of the landscape.
Prior to the period of slow-roll inflation 
that generates the density perturbations,
the volume distribution of inflationary regions with mass
parameters between $m$ and $m+dm$ and field values 
between $\phi_i$ and $\phi_i+d\phi_i$ is 
${\cal I}(m,\phi_i)dm \; d\phi_i$. 
Virtually nothing is known about the form of this distribution.  
Classical, slow-roll inflation occurs for field values in the 
range $M_{\rm pl} < \phi <M_{\rm pl}^{3/2}/m^{1/2}$. 
We do not consider the region with  $\phi \gg M_{\rm pl}^2/m$ where 
the vacuum energy density exceeds $M_{\rm pl}^4$, nor even
$\phi > M_{\rm pl}^{3/2}/m^{1/2}$ where the field evolution is 
quantum rather than classical \cite{Linde-eternal}.  Any period of inflation
governed by quantum evolution will have its effects included in the initial distribution
${\cal I}(m,\phi_i)$.
 
The epoch of chaotic inflation multiplies the initial volume 
 of each patch by an inflationary factor 
${\cal V} \propto e^{3N_e}$, with the e-fold number given by 
\begin{equation}  
 N_e \sim \frac{\phi_i^2}{M_{\rm pl}^2}.
\label{eq:Ne-chaotic}
\end{equation}
After reheating, local patches undergo power-law expansion 
until cold dark matter dominates the universe, structure 
begins to form, and hydrogen stars begin to shine. 
Such power-law expansion of the volume certainly has $m$ dependence, 
through the reheating temperature for example, but this is a negligible 
effect compared with the exponential increase of volume during inflation. 
Thus, the final physical volume 
$d V_{\rm phys} \equiv {\cal I}(\xi) {\cal V}(\xi) d \xi$ of patches with
initial inflationary parameters $(m,\phi_i)$ is roughly 
\begin{equation}
 dV_{\rm phys} \sim 
  e^{3 N_e (\phi_i)} \; {\cal I}(m,\phi_i) \; dm \, d\phi_i. 
\end{equation}

Density perturbations are generated by quantum fluctuations of the
inflaton and have a magnitude 
\begin{equation}
 \sigma \sim \frac{m \phi_{eq}^2}{M_{\rm pl}^3}.
\label{eq:delta1}
\end{equation}
The density perturbations that we observe were created when the field
value during inflation was $\phi_{eq}$, given by
$\phi_{eq}^2/M_{\rm pl}^2  \approx  \ln (T_{RH}/T_{eq})$. 
We assume instant reheating after inflation to a temperature $T_{RH}$.
In evaluating $\sigma$ from (\ref{eq:delta1}) to leading order, 
we ignore the temperature logarithm and take $\phi_{eq} \approx M_{\rm
pl}$, giving 
\begin{equation}
\sigma \sim m/M_{\rm pl}.
\label{eq:delta2}
\end{equation}
Since $\sigma$ depends only on $m$, and the  parameter $\phi_i$ cannot 
be measured, we can obtain the volume distribution for the observable 
$\sigma$ by integrating over $ \phi_i$:
\begin{equation}
 \frac{d V_{\rm phys.}}{d \sigma} \propto
  \int_{\phi_i^{ \rm min}}^{\phi_i^{ \rm max}} 
  e^{3 \phi_i^2/M_{\rm pl}^2} \; {\cal I}(m,\phi_i) \; d \phi_i, 
\label{eq:prob-chaotic}
\end{equation}
with $\phi_i^{\rm min} \sim \phi_{eq}$, and
\begin{equation}
\phi_i^{\rm max} \sim M_{\rm pl}^{3/2}/m^{1/2} \sim M_{\rm pl}/\sigma^{1/2}.
\label{eq:phimax}
\end{equation}

If the initial distribution ${\cal I}(m, \phi_i)$
has a milder dependence on parameters
than the volume factor $e^{3N_e}$ (we relax this assumption later),
the integration over $\phi_i$ is dominated by $\phi_i^{ \rm max}$. 
\begin{figure}
\begin{center}
\begin{picture}(230,195)(0,0)
 \Text(215,18)[lt]{$\phi_i$}
 \Text(37,182)[rb]{$\frac{m}{M_{\rm pl}} \approx \sigma$}
 \Text(28,117)[rb]{$\sigma \approx 10^{-4}$}
 \Text(28,83)[rt]{$\sigma \approx 10^{-6}$}
 \Text(116,150)[l]{$\phi_i^{max} \approx M_{\rm pl}^{3/2}/m^{1/2}$}
 \Text(79,145)[r]{$\phi_i^{min}$}
 \resizebox{7.5cm}{!}{\includegraphics{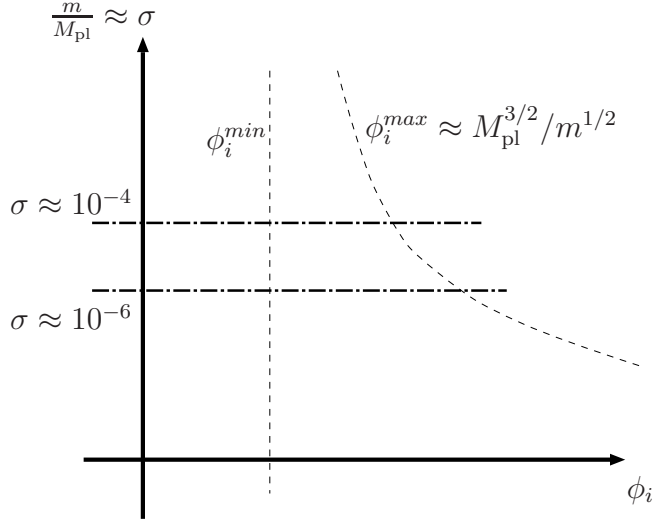}} 
\end{picture}
\end{center}
\caption{Parameter space of the chaotic inflation landscape. 
The density perturbation $\sigma$ depends on only one parameter 
of this model, $m$, while the volume increase due to slow-roll inflation 
is determined by $\phi_i$.  $N_e$ and $\sigma$ 
are related by an $m$-dependent upper bound on $\phi_i$.
}
\label{fig:chaotic}
\end{figure}
The probability distribution is then approximately given by:
\begin{equation}
 \frac{d {\cal P}(\sigma)}{d \sigma} \propto e^{1/\sigma}
 {\cal A}(\sigma,\Lambda^4).
\label{eq:exp-chaotic}
\end{equation}
Since the anthropic factor ${\cal A}$ does not depend too strongly 
on $\sigma$ for $10^{-6} \simlt \sigma \simlt 10^{-4}$, 
the volume factor ${\cal V}(\sigma) \propto e^{1/\sigma}$ 
will dominate the ${\cal P}(\sigma)$ distribution, 
making $\sigma$ as small as possible.  
This implies that an exponentially small faction of observers in the universe
see $\sigma \sim 10^{-5}$ in the middle of the anthropically allowed window;  
for example,
\begin{equation}
{\cal P}(10^{-5} < \sigma <  10^{-4})
       \approx e^{-10^{6}} \times {\cal P}(10^{-6} < \sigma < 10^{-4}).     
\label{eq:small-prob}
\end{equation}
This clearly indicates that either some of the assumptions about 
the underlying landscape are wrong, or we are far from being generic observers 
in the universe.

It is interesting to note that the edges of the anthropic window for $\sigma$ 
are not hard. 
If $\sigma$ is less than $10^{-6}$, the probability for a mode $\delta$
corresponding to typical large scale structures to fluctuate up to $10^{-6}$, 
as required for acceptable structure formation, is roughly $e^{-(10^{-6}/\sigma)^2}$.
Hence, with a flat initial distribution, 
${\cal P} \propto {\cal V} {\cal A} \sim e^{1/\sigma} e^{-(10^{-6}/\sigma)^2}$. 
If more sub-structures are necessary in galaxies for anthropic reasons, 
and if more seeds of the density perturbations are necessary for that
purpose, the anthropic factor may decrease faster than 
${\cal A}\approx e^{-(10^{-6}/\sigma)^2}$ as $\sigma$ becomes smaller 
than $10^{-6}$. 
But as long as the anthropic conditions only require large enough density 
fluctuations for a moderate number of modes, the volume factor is so powerful 
that $\sigma$ is pushed to smaller values, very far from the 
``anthropic window''. This observation tells us that 
${\cal P}(10^{-6} < \sigma < 10^{-4})$ itself in (\ref{eq:small-prob}) 
is much smaller than 1.
The anthropic conditions require $\delta$ to arise from fluctuations further 
out on the exponential tail of the Gaussian distribution.

\subsection{Multi-Parameter Model--- Hybrid Inflation Ensemble}
\label{ssec:hybrid}

Above, we assumed the expansion $V = m^2 \phi^2$ for the inflaton potential 
about each relevant local minimum of the landscape, with inflation occuring 
for field values $\phi > M_{\rm{pl}}$. However, it is much more reasonable 
to assume that the potential contains a constant term
\begin{equation}
V =  M^4 + m^2 \phi^2,
\end{equation}
with inflation able to occur for field values less than $M_{\rm pl}$. 
Each patch now undergoes hybrid inflation.\footnote{
The potential in the waterfall direction is omitted here because it is 
irrelevant during the inflationary era. Various 
standard-model minima may be associated with different types of inflation 
models, such as new inflation, but, for simplicity, we consider only 
the ensemble of hybrid inflation models. 
The conclusion in this sub-section---that the volume factor tends to depend 
on $\sigma$ exponentially---remains the same when a more generic ensemble 
of inflation models is considered.}
We assume that cosmological scanning occurs for both parameters $M$ and $m$, 
and for the intial and final values of the inflaton,  $\phi_i$ and $\phi_f$.
This potential involves three free parameters, so that there is no longer
a one-to-one correspondence between the density perturbation 
$\sigma$ and the parameters in the inflaton potential.
As we will see below, however, when the volume factor ${\cal V}$ is obtained 
as a function of $\sigma$ by integrating over all unobservable parameters,
it is exponentially sensitive to $\sigma$ in this model as well.

The number of e-foldings from this potential is: 
\begin{equation}
N_e \sim \frac{M^4}{m^2 M_{\rm pl}^2} 
                     \ln \left( \frac{\phi_i}{\phi_f} \right).
\end{equation}
As long as $N_e \gg 1$, $\phi_{eq} \approx \phi_f$  and the density 
perturbations at the epoch of matter-radiation equality are of order
\begin{equation}
\sigma \sim \frac{M^6}{M_{\rm pl}^3 m^2 \phi_f}.
\end{equation}
Assuming that the parameters are scanned in the ranges $m \simlt M_{\rm pl}$, 
$M_{min} \simlt M \simlt M_{\rm pl}$ with a phenomenological lower limit 
on $M_{min}$ from reheating, and $\phi_f \simlt \phi_i \simlt M_{\rm pl}$, 
we find that the e-fold number becomes the largest for a given $\sigma$ 
in a patch with $\phi_i$ as large as possible,  $M,m$ both as small 
as possible and $\phi_f \sim M_{\rm pl}/e$. 
\begin{figure}[htb]
\begin{center}
\begin{picture}(300,180)(0,0)
\Text(291,32)[l]{$m$}
\Text(267,30)[rb]{$M_{\rm pl}$}
\Text(45,165)[l]{$M$}
\Text(35,144)[rb]{$M_{\rm pl}$}
\Text(36,61)[rb]{$M_{min}$}
\Text(89,95)[rb]{$(\sigma)$}
\Text(110,86)[lb]{$(N_e)$}
\Text(160,105)[rb]{$\sigma>1$}
\Text(177,90)[lt]{$N_e$ is too small}
\resizebox{10cm}{!}{\includegraphics{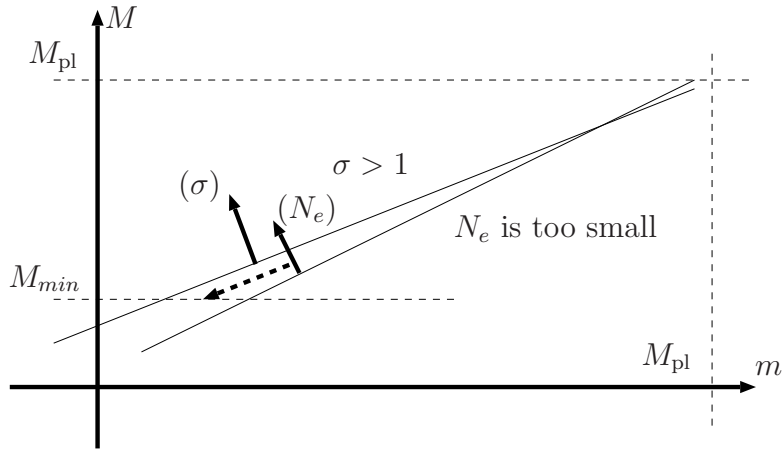}}
\end{picture}
\end{center}
\caption{Schematic parameter space of hybrid inflation models for fixed $\phi_i$ and $\phi_f$. 
Sufficient e-folding is not obtained in the lower-right region, while 
the density perturbation is too large in the upper-left region. 
 Directions normal to the contours of $\sigma$ and $N_e$ are 
indicated by two arrows in the figure, and are slightly different. 
Thus, on a contour of $\sigma$, the e-folding number $N_e$ increases 
in the direction shown by the broken arrow. For a given $\sigma$, 
$(M(\sigma),m(\sigma))$ on the line $M=M_{min}$ provides the largest 
e-folding number $N_e(M(\sigma),m(\sigma))$. With $\phi_i \sim M_{\rm pl}$ and 
$\phi_f \sim M_{\rm pl}/e$ the volume increase 
factor is roughly given by $e^{3N_e(M(\sigma),m(\sigma))}$.}
\label{fig:hybrid}
\end{figure}
Assuming again that the initial distribution factor ${\cal I}(M,m,\phi_i)$ 
is less important than the volume factor, 
the physical volume is exponentially dependent on $\sigma$
\begin{equation}
 d V_{\rm phys.} \propto 
   e^ {\frac{3M_{\rm pl}^2}{M_{min}^2} \sigma} \;  d \sigma,
\label{eq:hybrid-vol}
\end{equation}
and strongly favors larger values of $\sigma$, in contrast to the case 
of the chaotic inflationary regions. 

As long as $\sigma \simlt 10^{-4}$, the anthropic factor $\cal {A}$ has 
only a mild dependence on $\sigma$, and thus the distribution 
of the physical volume essentially determines the total probability distribution.
As in the model in section \ref{ssec:chaotic}, a negligibly small
fraction of observers in the universe sees $\sigma$ close 
to what we observe.\footnote{
This result does not depend on the particular choice of the boundary 
of the parameter space $M_{min}\simlt M$.}
Although one might like to consider the possiblity of a landscape of vacua in order 
to solve problems such as that of the CC, the model landscape discussed here is clearly not an acceptable one.

As in the previous sub-section, the exponential volume factor is so powerful 
that the anthropic window is forced to open wider. 
In this hybrid inflation landscape a typical observer will measure 
$\sigma > 10^{-4}$. With such large density perturbations typical planets 
will have their orbits disrupted before observers can form \cite{RT}, 
but a few planets will by chance avoid close contact with foreign stars for 
a sufficent time for observers to form, and, given the huge increase 
in the volume of the patch from inflation, such observers will dominate.

\subsection{Generalization}

The above two examples demonstrate a rather generic feature of landscapes 
that can be approximated by an ensemble of slow-roll inflation models 
with scanned parameters: 
the probability distribution over $\sigma$ contains a volume factor 
${\cal V}(\xi)$ that depends exponentially on $\sigma$. 
Hence, barring an important effect from the initial volume distribution 
${\cal I}(\xi)$,
a negligibly small fraction of observers in the universe sees a scale-invariant 
density perturbation of order $10^{-5}$. Whether the typical $\sigma$ is larger 
or smaller than $10^{-5}$ depends on the ensemble of inflation models, 
but either way, the density perturbation is predicted incorrectly.
Below we give a generalized argument for this ``$\sigma$ problem.''

Consider any two sub-universes, $i$ and $j$, in which large scale structure 
forms. We assume that all such sub-universes underwent a period 
of slow-roll inflation with a collection of parameters and fields, $\xi$, 
such as $(m,\phi_i)$ or $(M,m,\phi_i, \phi_f)$, that scan from 
one patch to another. We assume that it is meaningful to discuss 
the relative probability for these two sub-universes\footnote{
There is a subtlety when infinite numbers of observers have to be accounted for 
in the probabilites (see e.g., \cite{GL,Guth,Tegmark}). 
One can try to deal with this problem by regularizing the infinites and 
taking a limit.  Here, we just assume that there is a meaningful definition 
of relative probability, and we do not specify what it is. 
Our conclusion should not be affected by the definition, unless a really specific choice is made.}
\begin{equation}
\frac{{\cal P}(\xi_i)}{{\cal P}(\xi_j)} = 
 \frac{{\cal I}(\xi_i)}{{\cal I}(\xi_j)} 
 \frac{{\cal V}(\xi_i)}{{\cal V}(\xi_j)}\frac{{\cal A}(\xi_i)}{{\cal A}(\xi_j)}.
\label{eq:relprob}
\end{equation}
Again, ${\cal I}$ is an initial condition factor, giving the volume distribution 
prior to slow-roll inflation, ${\cal V}$ is the volume expansion factor from 
slow-roll inflation, generally having exponential dependence on $\xi$, 
and ${\cal A}$ contains all other anthropic factors
including those that prefer $\sigma$ to lie within the window 
$10^{-6} \simlt \sigma \simlt 10^{-4}$.  

If the density perturbations arise from quantum 
fluctuations of the inflaton field, then their standard deviations 
are determined by $\xi$.
The relative probability of finding two different values 
of the density perturbations, $\sigma_1$ and $\sigma_2$, will have the form 
\begin{equation}
 \frac{{\cal P}(\sigma_1)}{{\cal P}(\sigma_2)}= 
 \frac{\sum_{\xi|\sigma=\sigma_1} {\cal I}(\xi){\cal V}(\xi) A(\xi)}
      {\sum_{\xi|\sigma=\sigma_2} {\cal I}(\xi){\cal V}(\xi) A(\xi)}, 
\end{equation}
where we sum over all sub-universes giving a particular value for $\sigma$.

As a result of the exponential dependence on $\xi$, such sums
will generally be dominated by a particular value of $\xi$, i.e., 
$\sum_{\xi|\sigma=\sigma_1}{\cal I}(\xi){\cal V}(\xi) {\cal A}(\xi) \sim 
{\cal I}(\xi^*(\sigma_1)) {\cal V}(\xi^*(\sigma_1)) {\cal A}(\xi^*(\sigma_1))$, 
where $\xi^*(\sigma_1)$ is the exponentially most probable value of $\xi$ 
with $\sigma=\sigma_1$. 
That is to say, even though many different sub-universes will generally 
have a given value of $\sigma$, we expect one of them to be 
exponentially more probable than all the others. 
Then the probability of finding a given $\sigma$ is roughly the same 
as the probability of finding that specific sub-universe, i.e.,
\begin{equation}
 \frac{{\cal P}(\sigma_1)}{{\cal P}(\sigma_2)}\approx 
 \frac{{\cal I}(\xi^*(\sigma_1)) {\cal V}(\xi^*(\sigma_1)) A(\xi^*(\sigma_1))}
      {{\cal I}(\xi^*(\sigma_2)) {\cal V}(\xi^*(\sigma_2)) A(\xi^*(\sigma_2))}.
\end{equation}
Within the window  $10^{-6} \simlt \sigma \simlt 10^{-4}$, where ${\cal A}$ 
has mild dependence on $\sigma$, we therefore 
expect exponential sensitivity in the probability 
distribution, arising from the volume factor ${\cal V}$. 
This exponential sensitivity persists even if ${\cal I}(\xi)$ has a strong 
dependence on $\xi$, either steeply increasing or falling with $\sigma$ across 
the anthropic window. In these cases, to demonstrate a $\sigma$ problem
we do not need to calculate the form of ${\cal I}$, we need only assume that 
it does not have an exponential dependence on $\sigma$ that precisely cancels 
that of ${\cal V}$. Since ${\cal I}$ and ${\cal V}$ have entirely different 
physical origins, such cancelling exponentials could only be accidental.

Although we have discussed so far only the case in which the density 
perturbations are generated by inflaton fluctuations, 
it is straightforward to extend the argument to the scenario 
in which they originate from fluctuations of another light field, $s$ 
\cite{LW,MT,DGZK}. 
Analogous arguments lead generically to a distribution 
$d V_{\rm phys}(H) \propto e^{f(H)} dH$, with exponential dependence 
on the Hubble parameter $H$.
The density perturbation is given by $\sigma \sim H/\vev{s}$, 
and the probability distribution for $\sigma$ is given 
by convoluting those of $H$ and $\vev{s}$, 
and is generically exponentially sensitive to $\sigma$. 

\section{Dominant Selection from ${\cal I}$}
\label{sec:eternal}

In this section we consider a special form for the initial volume 
distribution that avoids the $\sigma$ problem. We take ${\cal I}$ to have 
a sharp peak within the anthropic window for $\sigma$ --- a peak that 
is so sharp that ${\cal IV}$ is also sharply peaked, so that the exponential 
behavior of ${\cal V}$ is sub-dominant. Indeed it may be that the 
discreteness of the vacua is relevant, so that a single anthropically acceptable 
vacuum has an initial volume very much larger than all the others.\footnote{
As before, we do not consider vacua that are anthropically unacceptable, say, 
because the QED fine-structure constant is outside its anthropic window, 
or because the CC has already dominated the universe 
by the epoch of recombination. 
These are excluded from consideration because there are so few observers 
in such vacua, and the associated anthropic factors ${\cal A}$ are essentially zero.}
Virtually all the observers in the universe, including us, will then see 
the physics of this vacuum, hopefully with $\sigma \sim 10^{-5}$ 
and $\Lambda^4 \sim (3 \times 10^{-3} \, \EV)^4$.

What kind of physics can prepare such an initial condition, and how 
probable is it that such a uniquely chosen vacuum happens to be ours?
In sub-sections 3.1 and 3.2,  we argue that eternal inflation can lead to 
such a strong selection of vacua; if the conditions for
eternal inflation are met in some patch of the universe, the enormous 
volume factor that results will play an important role in determining the 
distribution of observers in that patch. Although we present eternal
 inflation as a possible example mechanism,
we stress that our essential conclusions in sub-section 3.3 on 
probabilities
depend only on the assumptions made for ${\cal I}$, 
and not on any particular mechanism for obtaining  ${\cal I}$.

\subsection{False-Vacuum Eternal Inflation}
\label{ssec:false}

We consider a landscape of vacua with differing energies, 
as required for an anthropic solution to the CC problem.
Regions of the universe in local minima with positive vacuum energy 
expand exponentially, while regions in local minima with large negative 
vacuum energy shrink to cosmological singularities. 
We are not interested in the latter regions, because no observers 
live there. 
A positive vacuum energy region, on the other hand, continually 
nucleates bubbles of vacua with lower energy \cite{ColemanLuccia}, 
while inflating with its associated Hubble parameter, $H$.  
This process of inflation and bubble nucleation continues forever 
if $H_{\rm eff.}\equiv H-\Gamma_{\rm tot}/H^3$ is positive, 
where $\Gamma_{\rm tot}$ is the sum of all the bubble nucleation rates.
This sort of eternal inflation may be a generic feature of 
landscapes \cite{Susskind,Dine,Susskind05}.

There may be more than one eternally inflating local minimum, 
each with its own effective expansion rate. 
As you look infinitely far into the future, however, the false vacuum
with the largest $H_{\rm eff.}$ will be arbitrarily larger 
in physical volume than all the others, and will dominate the universe
\cite{Vilenkin95,GL}.\footnote{The physical volume of each false vacuum
depends on the choice of equal-time surface.
If inflation ends within a finite time, this subtlety is not 
a problem. But, since the false vacua are inflating forever, 
it is quite subtle to compare the two infinite numbers of observers 
produced in the bubbles nucleated from two different false vacua. 
For more about this issue, see \cite{GL,Guth,Tegmark}.
} This will be true even if its expansion rate is only very slightly larger 
than all the others.
This feature is convenient  because it causes any prior initial condition 
of the universe,  such as those  found in \cite{Vilenkin82,HH,Linde-wavefcn,
HT,Cornell}, to be erased \cite{Linde-eternal}.  
The universe converges to a fixed asymptotic state, dominated by a single 
local minimum and its associated bubbles.

How does this asymptotic-state universe\footnote{The dominant 
percentage of observers live in bubbles that nucleated at later times. 
This is why any local minima other than the asymptotic-state local minimum are 
irrelevant.} prepare an environment in which observers can live?
Some of the nucleated bubbles will go to standard-model vacua\footnote{
The decay to standard-model vacua includes cascade decays 
through various other vacua in
intermediate steps. What we call the asymptotic-state local minimum 
is assumed to have a non-zero decay rate to a standard-model vacuum. 
If it does not, it is replaced by the one with largest $H_{\rm eff.}$ 
among those that have non-zero decay rates to standard-model vacua. }  
with moderate values of the CC, and also 
with anthropically acceptable values for other parameters.
Note, however, that simple bubble nucleation to a standard-model vacuum 
is not sufficient to create a habitable universe. 
The space inside the bubble must be reheated, and furthermore, must 
become a flat universe, rather than an open universe.\footnote{This may 
be an anthropic condition because density perturbations do not grow 
in curvature dominated backgrounds \cite{KT}.  If it is not, however, 
we just assume that the bubble with the largest $\Gamma_i$ (see what 
follows in the text) happens to go to a slow-roll region. 
More discussion on this issue is found in \cite{Susskind05} and 
    references therein.} 
These conditions are most readily satisfied if a nucleated bubble 
goes not directly to a standard-model minimum, but rather to a slow-roll 
inflationary region that reheats to a standard-model minimum. 
For this reason we neglect bubbles that do not nucleate to slow-roll 
inflationary regions. The result is that false-vacuum eternal inflation 
sets the initial volume distribution ${\cal I}(\xi)$ for  the slow-roll 
inflation ensemble.

What sort of initial volume distribution is obtained?
Let $\Gamma_i$ be the bubble nucleation rate from the dominant local 
minimum to a given inflationary region, labeled by $i$;
 the nucleation rates to two different such  regions of the landscape will generally 
have different values.  Now, the total volume of all sub-universes produced in the region $i$ will simply
 be proportional 
to $\Gamma_i$, with the physical volume of the dominant eternally 
inflating local minimum factoring out; ${\cal I}(\xi_i) \propto \Gamma_i$.
An individual decay rate $\Gamma_i$ takes the form $M_i^4 e^{-S_i}$, 
where $M_i$ is a characteristic energy scale of the potential barrier 
and the distance of the tunneling, and $S_i$ is the classical action 
of a bounce solution interpolating between the two vacua.   Now note that
 ${\cal I}(\xi_i)$ cannot be expected to vary mildly
as a function of the low-energy parameters $\xi_i$; 
while two slow-roll inflationary regions neighboring each other in a landscape may have 
much the same tunneling rate $\Gamma_i$'s, their low energy parameters, 
such as the inflaton mass and the CC, may be totally different, 
as in the case where the CC is given by the mechanism found in \cite{BP}.
 Two inflationary regions with almost 
the same low-energy parameters may generally be far away from each other in the landscape of vacua, 
with $e^{-S_i}$ factors differing by a huge amount.

If the landscape does not have large numbers of anthropically acceptable standard-model 
vacua, then it will not be possible to treat
the initial volume distribution ${\cal I}(\xi)$ as 
a continuous mild distribution over the low-energy parameters, even after 
binning and averaging.  It will rather become an essentially isolated distribution with perhaps a few  
exponentially high peaks.
We assume that this exponential dependence from $e^{-S_i}$ is 
more important than the volume increase from slow-roll inflation. 
This is how eternal inflation might be able to prepare the sort of initial volume distribution proposed
at the beginning of this section.
Perhaps the standard-model vacuum in closest proximity to the dominant 
eternally inflating vacuum will be almost uniquely selected.

Essentially unique values are thus chosen 
for various parameters in this type of scenario, including $\sigma$ and 
$\Lambda^4$, in the sense that the same values are observed by virtually all observers 
in the universe. They therefore should clearly be the values that {\it we} observe. 
We cannot presently test this idea, since we do not at this moment have a guess as to the
 details of the underlying landscape. 
One could in principle work out which is the 
dominant eternally inflating local minimum in a given landscape, 
which inflationary region has the largest bubble nucleation rate from 
the minimum, and what is the value of the CC 
for all the relevant standard-model vacua.
This might be doable once a concrete landscape, such as 
the Type IIB string landscape, is adopted.

\subsection{Large-Field Eternal Inflation}
\label{ssec:large-field}

Eternal inflation can also take place by another  mechanism 
\cite{Linde-eternal}.
When a slow-roll inflaton potential is so flat that the condition
\begin{equation}
 H \simgt \frac{\dot{\phi}}{H} \approx \frac{V'}{H^2}
\label{eq:eternal-cond}
\end{equation}
is satisfied, the evolution of the inflaton is mostly governed by 
quantum fluctuations, and not by the classical equations of motion. 
If this is the case, the average value of the inflaton field, weighted by 
the physical volume, does not descend the potential, but goes uphill, 
because the expansion rate of the volume is higher for a larger 
energy density \cite{Linde-eternal,uphill}. One such eternally 
inflating region eventually dominates the volume of the universe: 
the one with the highest expansion rate \cite{GL}, 
just as in the false-vacuum eternal inflation case. 

The quantum fluctuations of the inflaton occasionally bring its value 
outside of the range satisfying (\ref{eq:eternal-cond}), 
converting some part of the eternally inflating spacetime 
into a classical slow-roll inflation ``bubble''.  
Once such a bubble enters a stage of slow-roll inflation, it is eventually 
reheated and leads to the standard cosmology. 
The bubble nucleation process in the false-vacuum type eternal inflation 
is replaced by the creation of quantum fluctuation bubbles in this
scenario. Note that eternal inflation must be followed by a period 
of standard slow-roll inflation in this scenario as well; during 
the eternal inflation density fluctuations are generated which are of order 
$\sigma \approx H^2/\dot{\phi}$; this is larger than 1 because of 
(\ref{eq:eternal-cond}) \cite{Linde-eternal}. 
When density fluctuations of order one enter the horizon, primordial 
black holes are produced, leading to a black-hole dominated universe 
\cite{PBH,RT}. Thus, the period of slow-roll inflation cannot be
skipped. 
Eternal inflation of large-field type thus also sets an initial condition 
${\cal I}(\xi)$ for the slow-roll inflation ensemble.  

Since the exiting process from the eternal inflation epoch 
is through quantum fluctuations, the history after the exit is not 
determined completely. 
There may be several paths from the dominant eternally inflating region 
to standard-model vacua, passing through slow-roll inflationary regions. 
The initial volume distribution ${\cal I}(\xi)$ is non-zero for 
such inflationary regions, and the relative ratio is calculated from the rates 
for quantum fluctuations to exit along these various routes. 
Rates for quantum fluctuations thus replace the bubble nucleation rates 
$\Gamma_i$ from the false vacuum eternal inflation case.
If only one path is favored significantly over the others, 
then one set of parameters is observed by almost everyone in the universe. 
All of the arguments based on false-vacuum eternal inflation thus hold
true in the large-field eternal inflation case.

\subsection{The Probability for Choosing Our Vacuum}
\label{ssec:large-field}

We now turn to the question of how likely it would be for the chosen value of $\sigma$ to be $10^{-5}$ 
and the chosen value of the CC to be $(3 \times 10^{-3} \EV)^4$.  We assume here  only that  the initial distribution
factor selects a particular anthropically acceptable vacuum as being the most probable one.  Specifically,
 we do not need to assume anything about eternal inflation in this sub-section. 
The analysis will be a bit subtle, but the basic idea is simple:  the larger the range of allowed values for
a parameter, the less likely it is that the particular value we see would be chosen.  Let us define a value to be ``choosable" 
 if any suppression in observers from the  anthropic factor ${\cal A}$ is less important than the increase
in the number of observers from the initial distribution ${\cal I}$.  Since the initial distribution factor is expected to be quite
strong, the range of ``choosable" values  for a parameter tends to be larger than one might expect from anthropic considerations alone.
The result is to make it {\it less} likely that we observe the values for $\sigma$ and the CC that we do.   
This may be a problem for this scenario.

For the moment, let us take it for granted that $\sigma$ 
is chosen correctly, and consider only the selection of the CC. We feel it to be a reasonable assumption
 that the initial distribution factor ${\cal I}$ 
doesn't have much dependence on the actual value of the very small CC that emerges
after reheating and the various phase transitions of late-time cosmology. The location of the vacuum 
within the landscape will be relevant to ${\cal I}$, 
but whether  its energy is $10^{-120} M_{\rm pl}^4$ or
  $10^{-121} M_{\rm pl}^4$ probably will not be.  We will then say that any anthropically acceptable 
vacuum has an ``equal chance" to be the one with the largest ${\cal I}$.  This is simply
a statement of our ignorance about the precise details of the underlying landscape.  Since 
the volume factor ${\cal V}$ is irrelevant in this picture, as explained above, the result is that for 
anthropically acceptable CC's, the probability for a specific value to be chosen is given
primarily by the fundamental density of states in the landscape.   This logic may apply to other parameters as well.
We will then assume that the density of states gives a flat distribution 
for small values of the CC, as in \cite{WeinbergI,WeinbergII}.
Since standard-model vacua with  CC's satisfying 
\begin{equation}
\Lambda^4 \simlt \left[\rho_{\rm CDM} \sigma^3 \right]_{rec}
\label{eq:WeinbergI-modified}
\end{equation}
are certainly satisfactory, at most only one part in a hundred 
anthropically acceptable standard-model vacua have a CC 
as small as ours: $P[\Lambda^4 < (3\times 10^{-3} \EV)^4] < 10^{-2}$.
Here we use $P$, rather than the ${\cal P}$ used earlier, to emphasize that 
this new probability is a statement of our ignorance about which vacuum happens to 
be selected, rather than a measure of a fraction of observers.

Those vacua satisfying (\ref{eq:WeinbergI-modified}), however, are not all 
the ``choosable" ones, in the sense defined at the beginning of this sub-section.
Since each slow-roll inflationary region is associated only with 
the standard deviation $\sigma$ of the density perturbations, 
there is a chance that the actual density fluctuations $\delta$ could be larger 
than $\sigma$, so that the true anthropic condition (\ref{eq:WeinbergI}) 
is satisfied even for a CC larger than (\ref{eq:WeinbergI-modified}).
If we adopt the estimate for the anthropic factor in \cite{WeinbergII}
\begin{equation}
{\cal A}(\Lambda^4) \approx 
 e^{-[(\Lambda^4/\rho_{\rm CDM})^{2/3}/\sigma^2]_{rec}} \qquad 
 {\rm for} \quad \Lambda^4 \simlt 10^{5} \times (3 \times 10^{-3} \EV)^4, 
\label{eq:A-for-large-Lambda}
\end{equation} 
with $\sigma_{rec}$(Mpc) of order a few times $10^{-3}$ corresponding 
to the COBE normalization \cite{WeinbergII}, 
we have ${\cal A} \sim e^{-100}$ for a CC $10^5$ times larger than ours.
Thus, the suppression of gaussian fluctuations ``slightly'' disfavors such a large CC, 
reducing the number of observers by a factor of $e^{-100}$. 
But this effect is not as significant as  that of the volume factor, 
which is expected to vary from one inflationary region to another 
by at least of order $e^{3N_e}>e^{100}$.
Since we have assumed in this section that the hierarchy among 
${\cal I}$ values is more than that among the volume factors, 
the anthropic factor is relatively negligible for  CC's $10^5$ times 
larger than ours, and perhaps larger. 
If we then consider vacua with $\Lambda^4 < 10^{5} \times (3\times 10^{-3} \EV)^4$
to be ``choosable", the probability $P[\Lambda^4 < (3\times 10^{-3} \EV)^4]$ is less than $10^{-5}$.
The anthropic factor ${\cal A}$ above, however, comes from the assumption 
that the density perturbation of a {\it single} wavenumber 
is required to go non-linear and form a massive clump before 
the CC dominates the energy density of the universe. 
This will certainly be a necessary condition for observers to exist, 
but may not be a sufficient condition \cite{WeinbergII}.
Thus the anthropic factor may decrease much faster than 
(\ref{eq:A-for-large-Lambda}), and it is not a certainty that the upper bound on the 
probability is less than $10^{-2}$.  

The lower bound for the probability is clearer, however;
 there is no chance for a reasonable scenario of 
structure formation when the CC is almost as large 
as the energy density at the epoch of recombination. 
Thus, vacua with $\Lambda^4>[\rho_{CDM}]_{rec}$ are not regarded as ``choosable" \cite{RT},
 even if there are ${\cal I}$ values for such vacua that are very large. 
We thus have 
\begin{equation}
10^{-9} \, \simlt \, P\left[\Lambda^4 < (3\times 10^{-3} \EV)^4 \right] 
        \, \simlt  \, 10^{-2},
\end{equation}
with $10^{-9}$ coming from the ratio $(3 \times 10^{-3} \EV)^4/[\rho_{CDM}]_{rec}$.

One could in principle ask a similar question about the probability that 
$\sigma$ would be chosen to be $\approx 10^{-5}$ in this scenario.  
The nature of the density of states distribution on $\sigma$, however, 
is hard to estimate, as $\sigma$ is a model dependent function of fundamental parameters 
of the landscape, and
therefore we do not attempt to calculate it. 
Even if it turned out that $\sigma \approx 10^{-5}$ was chosen, however, 
we would not have an explanation for why this chosen value is within the 
anthropically preferred window of $10^{-6} \simlt \sigma \simlt 10^{-4}$; 
the window is not a hard cut-off, just as the upper bound on the CC 
(\ref{eq:WeinbergI-modified}) is not.  The strength of the initial distribution factor can 
overcome suppression in the number of observers when $\sigma$ lies outside the window.

Now, the e-fold number $N_e$ of the last slow-roll inflation epoch is also 
an observable for values of about 60. Thus, the density of states 
can also be represented as a function of both $\sigma$ and $N_e$. 
Reference~\cite{Susskind05} discusses the density of states as a function of 
$N_e$, based on a simple model.
While that paper ignored volume factors in probabilities, the scenario 
outlined here could provide a justification for this approach.  One
 need only keep in mind that the resulting probabilities are statements of 
ignorance rather than distributions of observers.  Discussions in \cite{Susskind05}
and references therein concerning the lower multipoles of the CMB, for example,  would then
be applicable.

In summary, the $\sigma$ problem of section \ref{sec:ensemble}
could be solved by an extremely
sharp peak in the ${\cal I}(\sigma)$ distribution.  This could possibly be achieved
from eternal inflation.  
By selecting  roughly a single standard-model vacuum, and thus single values 
for both $\sigma$ and $\Lambda^4$, this mechanism could
circumvent the exponential dependence on $\sigma$ coming from 
slow-roll inflation.  
The biggest problems with this scenario however are
\begin{itemize} 
\item The probability 
$P(\Lambda^4 < (3 \times 10^{-3}\EV)^4)|_{\sigma = 2\times 10^{-5}}$ 
is at least as small as $10^{-2}$ and may even be as small as $10^{-9}$. 
The broad range comes from uncertainty in anthropic conditions as well as model dependence.
In any case this probability is worse than the 5--10\% of \cite{WeinbergII}, but better than $10^{-120}$.
\item  There is no reason for the observed spectrum of density perturbations 
to fall within the middle of the anthropically preferred window 
$10^{-6} \simlt \sigma \simlt 10^{-4}$.
\end{itemize}

\section{Preferred Landscapes Without a $\sigma$ Problem}
\label{sec:restricted}

In the previous section we have shown that the $\sigma$ problem may 
be solved if the dominant vacuum selection is determined by the initial 
condition factor ${\cal I}$, rather than by the volume factor ${\cal V}$, 
but at the cost of the two problems listed above.
In this section we seek alternative solutions to the $\sigma$ problem; 
in particular ones in which the physical volume distribution ${\cal IV}$, 
named the ``a priori factor'' in \cite{WeinbergII}, has a flat distribution 
in the CC, leading to the success 
${\cal P}(\Lambda \leq 3 \times 10^{-3} \EV)|_{\sigma = 2\times 10^{-5}} 
\sim $ 5--10\% for the CC problem \cite{WeinbergII}.
If ${\cal IV}$ is moderately peaked near $\sigma \sim 10^{-5}$, we have 
essentially the assumption made in \cite{WeinbergII}.
If ${\cal IV}$ is flat (or at most power-law) in $\sigma$ 
across the anthropic window as well ({\it c.f.} \cite{Wise}), 
the anthropic factor naturally accounts for why $\sigma_{\rm COBE}$ 
happens to lie within the anthropic window.
In both cases the key is to avoid an exponential behavior for 
${\cal V}(\sigma)$, the $\sigma$ problem  in section \ref{sec:ensemble}. 
In fact it seems that the scanning in the inflaton sector must be restricted 
in some way. 

The most obvious solution to the $\sigma$ problem is that none of the 
parameters of slow-roll inflation scan significantly. 
The parameters may be uniquely determined, or the density of states 
as a function of the parameters may have a sharply peaked behavior 
\cite{ADK,Acharya}. 
Since $\sigma$ is not scanned, the $\sigma$ problem does not exist.
Scanning of the CC can still occur, for example from the scanning 
of the parameters of the hidden sector that leads to supersymmetry breaking.
As long as ${\cal I}$ is roughly flat in the CC, the successful result of \cite{WeinbergII} 
follows.
This solution to both the $\sigma$ and CC problems, however, may not 
leave behind an anthropic explanation for the flatness of the inflaton potential. 

Another solution to the $\sigma$ problem results if the scanning of 
the inflaton sector is restricted in such a way that while $N_e$ and $\sigma$ 
scan they do not depend on a common 
scanning parameter of the theory. 
In this case $N_e$ is scanned cosmologically and the flatness problem of  
the inflaton potential is solved. 
Since the scanning of $N_e$ is independent of the scanning of $\sigma$, 
${\cal V}$ has no exponential sensitivity to $\sigma$. Thus, the $\sigma$ 
problem is avoided. As long as ${\cal I}$ is approximated by a mild function 
of $\Lambda^4$ and scanning parameters of slow-roll inflation, 
${\cal IV}$ varies mildly across the anthropic window of $\sigma$, and 
most of the observers in the universe are likely to see $\sigma$ in 
the middle of the anthropic window $10^{-6} \simlt \sigma \simlt 10^{-4}$.
The physical volume distribution ${\cal IV}$ may be flat in $\Lambda^4$ 
as above, and then it follows that 
${\cal P}(\Lambda < 3 \times 10^{-3} \EV)|_{\sigma = 2\times 10^{-5}} 
\approx $ 0.05--0.10 \cite{WeinbergII} or 
${\cal P}(\Lambda < 3 \times 10^{-3} \EV) \sim 10^{-4}$ as in \cite{Wise}.
Such restricted scanning is possible in both chaotic and hybrid inflation 
models, as shown below.

For the chaotic inflation ensemble, in contrast to sub-section 
\ref{ssec:chaotic}
we now assume that the quadratic potential extends only 
to field values somewhat larger than the Planck scale, say 
$10 \times M_{\rm pl}$, rather than to the very large $m$ dependent  
$\phi_i^{\rm{max}}$ of (\ref{eq:phimax}). 
In this case (\ref{eq:exp-chaotic}) is replaced by 
\begin{equation}
\frac{d \, {\cal P}}{d \, \sigma} \propto 
e^{100} {\cal I}(m,\phi_i^{max})|_{m\sim \sigma M_{\rm pl}} \cal{A}.
\end{equation}
Since the upper bound on the initial field value is no longer tied to 
the mass parameter or to the observed density perturbation, 
the volume increase factor from slow-roll inflation, $e^{3 N_e}$, 
is no longer an exponential function of $\sigma$.  
Actually, $\phi_i^{max}$ could have a moderate 
$m$ dependence, so long as such dependence did not lead to a 
dominant exponential distribution in $\sigma$.  

For the hybrid inflation ensemble of \ref{ssec:hybrid}, we now scan 
$m^2$ and the initial and final field values, but fix the parameter $M$. This 
restriction on the scanning might result if the model is extended to 
include  grand unification \cite{GUT-hybrid}, so that
$M$ is the unified symmetry breaking scale. Different values 
of $M$ lead to different QED and QCD coupling constants, so that the 
anthropic factor ${\cal A}$ very strongly selects a narrow range for $M$.
Since $m^2$ is scanned and so is $N_e$, the fine-tuning problem of the inflaton 
mass is solved anthropically \cite{Vilenkin95}. 
However, since both $N_e$ and $\sigma$ depend on $m^2$, which scans, 
we have not yet achieved our objective, which was to solve the $\sigma$ problem. 
The solution is for density perturbations to arise from
a mechanism such as those found in \cite{LW,MT,DGZK}, rather than 
from the inflaton fluctuation itself. 
The initial field value of another light field $\vev{s}$
gives rise to the density perturbation $\sigma \sim H/\vev{s}$, which is independent 
of $m^2$, as the Hubble parameter during inflation depends only on $M$ and 
does not scan.\footnote{We should clarify how it is possible to scan the CC 
but not the vacuum energy of slow-roll inflation.
The vacuum energy of inflation has two contributions: 
one from the CC and the other 
from the ``waterfall energy'' released at the end of inflation.  
Scanning of the CC piece certainly does affect 
the Hubble parameter during inflation, but it is anthropically constrained 
to be so small that its effect on the inflation Hubble parameter 
is negligible.}
Thus the volume of a patch is not correlated with its density
perturbation $\sigma$. The behavior of ${\cal I}$ across the anthropic window 
depends on the distribution of the initial value for $\vev{s}$, and could be 
flat or power-law in $\vev{s}$. 
Then the likely value of $\sigma$ would be determined mainly by the anthropic 
factor. 

A crucial assumption made in these solutions is that ${\cal I}$ can be treated 
as a mild distribution in $\Lambda^4$ and the scanning parameters of slow-roll inflation.
Could this assumption still be reasonable 
if eternal inflation occurs?
For false-vacuum eternal inflation, the previous 
section used the picture that the exponential dependence coming from 
$\Gamma_i \propto e^{-S_i}$ was so large that ${\cal I}$ was very sharply peaked
at particular vacua.  This would not necessarily be the case, however.  If there were large numbers of 
anthropically acceptable standard-model vacua 
in the landscape, or if there were clumps of such vacua in close proximity to each
 other resulting in similar $\Gamma_i$ values, 
then it might be possible for ${\cal I}$ to be a mild distribution even in the presence of eternal inflation.

\section{Conclusions and Discussion}

We have considered models of landscapes motivated by the CC problem and 
by the severe anthropic constraints on various other parameters 
of the standard model. If the parameters of a slow-roll inflation 
model are scanned cosmologically, this may further explain the existence of a flat inflaton potential;
although very flat potentials may be rare within the landscape, the exponential increase in volume 
that results would more than make up for this rarity.  A large volume factor leads to a large number of observers
to see it.

On the other hand, the volume factor discussed here is likely to be so powerful in determining the observed density 
perturbations, that its effect should be carefully studied. 
When a landscape of vacua is approximated by an ensemble of slow-roll 
inflationary regions, we find that the volume factor is generically 
exponentially sensitive to the density perturbation $\sigma$, so that an 
{\it exponentially} small fraction of observers in the universe see $\sigma$
of order $10^{-5}$.
Hence, such landscapes do not provide a viable setting for understanding the value of 
the small but non-zero CC.

Two ideas to avoid the $\sigma$ problem have been presented in 
this article. There may well be others. In one of them, we assume an initial volume distribution 
${\cal I}$ that gives an exceedingly large weight to one of the anthropically acceptable standard-model 
vacua.  In particular, this weight was assumed 
to be much larger than the other factors, so that the volume factor was relatively 
unimportant and the $\sigma$ problem was absent. 
Virtually all observers in the universe, including ourselves, 
see the same values for the low-energy parameters in this scenario.  
The values for parameters that would actually be chosen cannot be identified, however,
unless a particular landscape of vacua is specified, and the resulting ${\cal I}$ determined. 
Replacing our ignorance of the landscape with a density of states that is 
independent of the CC, the probability of the CC being chosen to be smaller than what we see 
is in the range $10^{-9} \simlt P[\Lambda^4<(3\times 10^{-3}\EV)^4] \simlt 10^{-2}$, with uncertainties
from anthropic conditions and model dependence.
Here we have just assumed that the density perturbation was chosen to be $10^{-5}$,
and we stress that there is no understanding of why $\sigma$ lies 
in the middle of its ``anthropic window'' in this scenario.
The key point is that the strength of ${\cal I}$ widens the anthropically allowed
ranges for parameters in this picture, reducing the probabilities that we measure {\it our} values.
Eternal inflation may have occured prior to slow-roll inflation, 
and could provide a dynamical mechanism for obtaining the sharply peaked 
${\cal I}$ assumed here.

In our second idea, we consider landscapes where the physical volume  
distribution ${\cal IV}$ does not have a $\sigma$ problem to begin with and in which
the distribution on the CC is roughly flat. 
If the e-folding number $N_e$ and $\sigma$ do not depend on a common 
scanning parameter of the inflation model, then the volume factor ${\cal V}$ 
does not necessarily depend exponentially on $\sigma$, 
solving the $\sigma$ problem. 
The naturalness problem for the inflaton mass is also solved through 
the scanning of $N_e$.
If ${\cal I}$ then has sufficiently mild dependence on parameters,
most of the observers in the universe will then see $\sigma$ to be within the 
``anthropic window'' $10^{-6} \simlt \sigma \simlt 10^{-4}$, and furthermore, 
the successful anthropic explanation for the CC will be in full form, with 
${\cal P}[\Lambda^4 < (3 \times 10^{-3} \EV)^4] \sim 5 - 10 \%$.

\section*{Acknowledgments}  

This work was supported in part by the Director, Office of Science, Office 
of High Energy and Nuclear Physics, of the US Department of Energy under 
Contract DE-AC03-76SF00098 and DE-FG03-91ER-40676, and in part by the 
National Science Foundation under grant PHY-00-98840.
We thank Matthew Schwartz for discussion.
T.W. thanks the Miller Institute for the Basic Research in Science.

\end{document}